\documentclass[11pt]{article}
\usepackage{moriond}
\usepackage{epsfig}
\usepackage{graphicx}
\usepackage{amssymb}
\usepackage{epstopdf}
\usepackage{graphicx}
\usepackage{geometry}    

\def\Journal#1#2#3#4{{#1} {\bf #2}, #3 (#4)}

\def\PLB{{\em Phys. Lett.}  B}

\def\PRD{{\em Phys. Rev.} D}

\def\be{\begin{equation}}
\def\ee{\end{equation}}
\def\bea{\begin{eqnarray}}
\def\eea{\end{eqnarray}}

\begin{document}
\vspace*{4cm}
\title{Measurement of the Underlying Event and Minimum Bias at LHC}

\author{ F. Ambroglini }

\address{University of Perugia, Department of Physics and INFN, via Elce di Sotto 1, 06123 PG, Italy.}

\maketitle\abstracts{
A study of \textit{Underlying Events} (UE) and \textit{Minimum Bias} (MB) at \textit{Large Hadron Collider} (LHC) with CMS and ATLAS detector under nominal conditions is discussed. Using charged particle and charged particle jets, it will be possible to discriminate between various QCD models with different multiple parton interaction schemes, which correctly reproduce Tevatron data but give different 
predictions when extrapolated to the LHC energy. This will permit improving 
and tuning Monte Carlo models at LHC start-up, and opens prospects for exploring 
QCD dynamics in proton-proton collisions at 14TeV.}

\section{Introduction}

One of the first physics results from the LHC will be the measurement of charged hadron spectra in proton-proton collisions and the estimation of the UE activities, a particularly interesting result given that interactions of protons have never before been observed in the laboratory at such high energies ($\sqrt{s}$ = 14 TeV). 
The measurement of these observable will also serve as an important tool for the calibration and understanding of the detectors and will help establishing a solid basis for exclusive physics. 
The ability to properly identify and calculate the UE activity, and in particular the contribution 
from \textit{Multiple Parton Interactions} (MPI)~\cite{mpi_ref} , has direct implications for other measurements at the LHC.

\section{Minimum Bias}

\subsection{Trigger strategy}

The minimum bias trigger has been designed in both the experiment ATLAS and CMS with similar approach. The CMS MB trigger~\cite{cmsmbtrigger} is based on counting towers with energy above the detector noise level, in both forward hadronic calorimeters (HF, 3$< |\eta| <$5). A minimal number of hits (1, 2 or 3) will be required on one or on both sides, an energy threshold value of 1.4 GeV is used in the hit definition. Once the luminosity is high enough, events can also be taken with the so called zero-bias trigger: a random (clock) trigger. The ATLAS trigger is based on MBTS~\cite{mbts} (\textit{Minimum Bias Trigger Scintillator}) detector mounted on the front face of the end-cap cryostats covering same radii as the inner detector 
(2$< |\eta| <$3.8). In order to fit with the expected rate the accidental rate from noise must be 
suppressed by a factor of $10^{7}$.

\subsection{Measurement}

\begin{figure}[htbp]
\centering
\includegraphics[scale=.75]{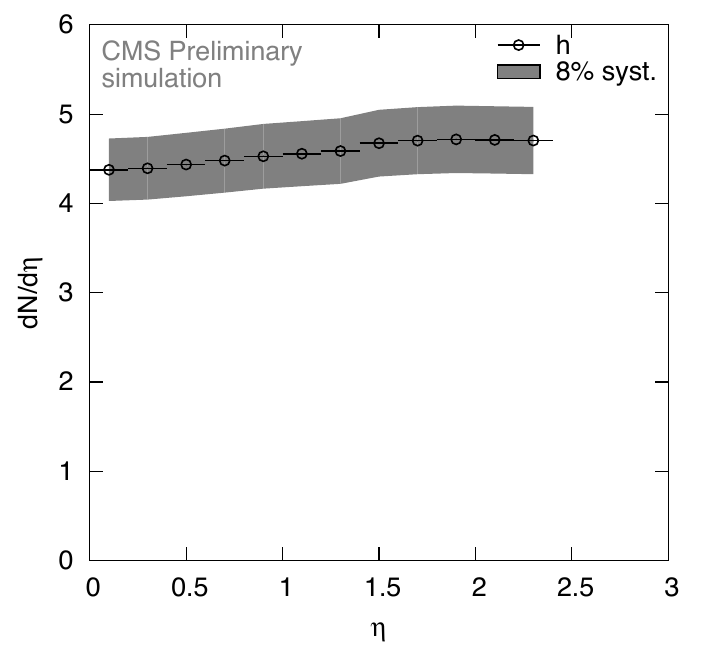} 
\includegraphics[scale=.75]{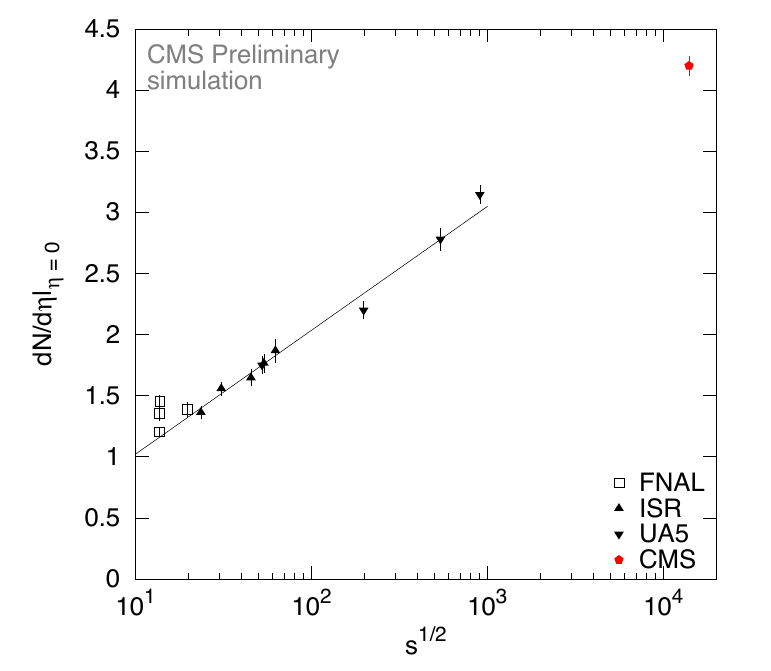} 
\caption{ CMS. Left: Pseudorapidity distribution of charged hadrons. Right: Energy dependence of pseudorapidity density of charged hadrons at $\eta = 0$.}
\label{fig:cms_mb}
\end{figure}

A good measurement of differential and integrated yields requires particle tracking down to as low $p_{T}$ values as possible. With a modified algorithm the pixel detector can be employed for the reconstruction of very low $p_{T}$ charged particles. The acceptance of the method extends down to 0.1, 0.2 and 0.3 GeV/c in $p_{T}$ for pions, kaons and protons, respectively. Charged particles can be singly identified or their yields can be extracted using deposited energy in the pixel and strip silicon tracker. %The invariant yields were fitted by the Tsallis function~\cite{tsallis}. 
The pseudorapidity distribution of charged hadrons and the energy dependence of some measured quantities can be studied (Figure~\ref{fig:cms_mb}).
\section{Underlying Event}

\begin{figure}[htbp]
\centering
\includegraphics[scale=.95]{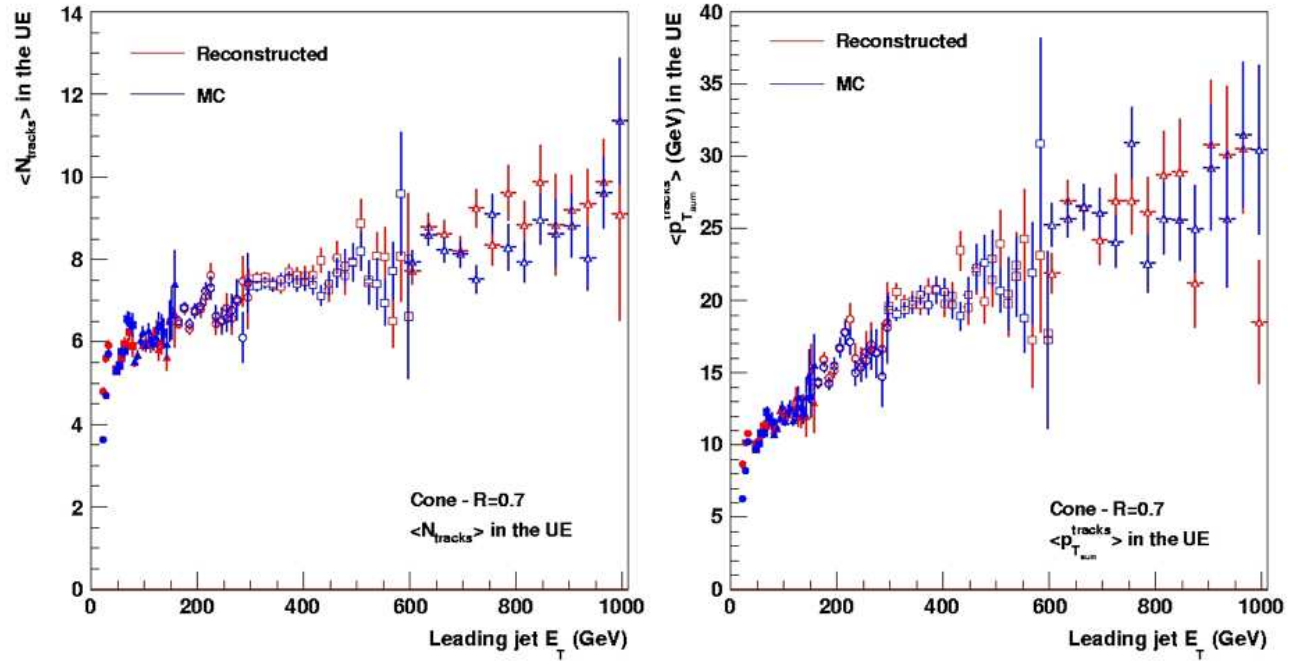} 
\caption{ ATLAS. Mean number of tracks (left) and track $p_{T}$ sum (right) for tracks with $p_{t} >1GeV/c$ and with $|\eta| < 2.5$, in the transverse region for both reconstructed (red) and MC (blue) versus the transverse energy of the leading calorimetric jet. }
\label{fig:atlas_ue}
\end{figure}

\begin{figure}[htbp]
\centering
\includegraphics[scale=.85]{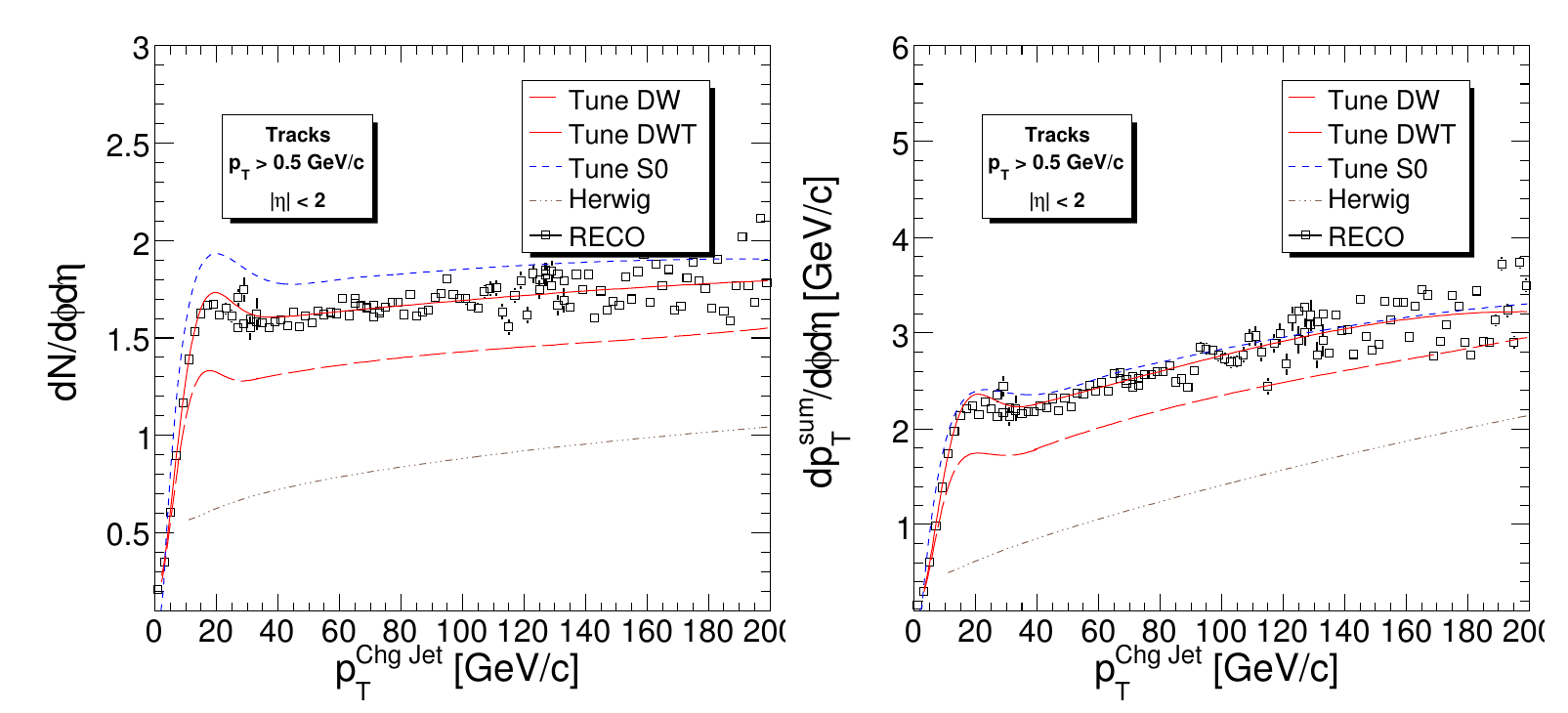} 
\caption{ CMS. Densities $dN/d\eta d\phi$(left)and $dp^{sum}_{T} /d\eta d\phi$ (right) for tracks with $p_{t} >0.5GeV/c$, as a function of the leading charged jet $p_{T}$, in the transverse region, for an integrated luminosity of 100 $pb^{-1}$ (corrected distributions).}
\label{fig:cms_ue}
\end{figure}

\begin{figure}[htbp]
\centering
\includegraphics[scale=1.]{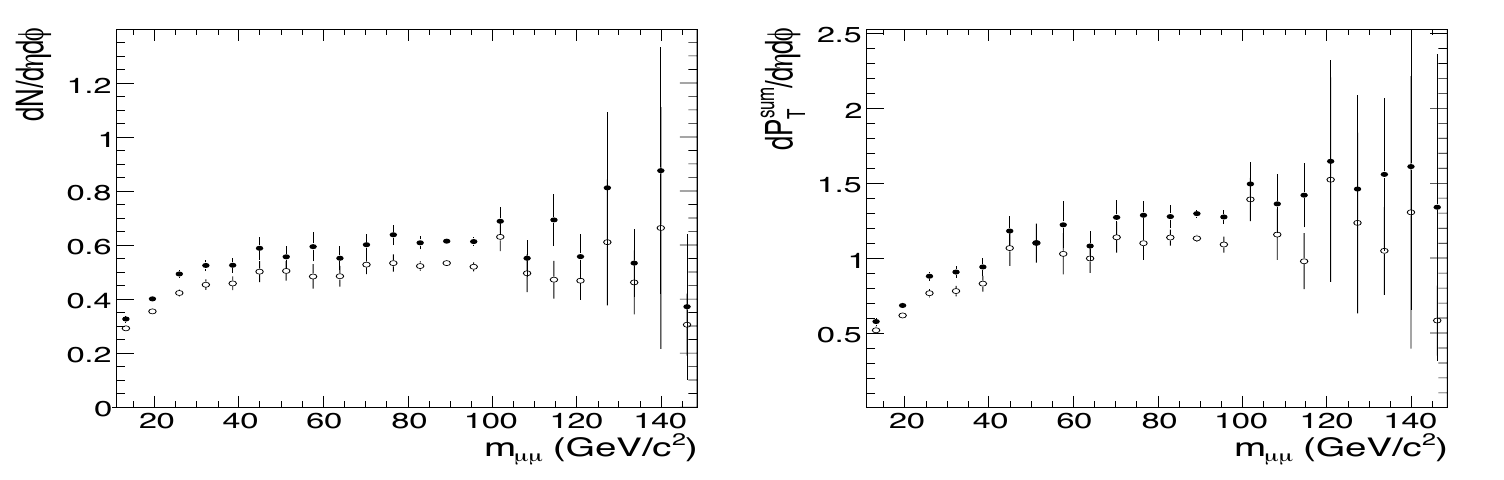} 
\caption{ CMS. Density of charged particles,$dN/d\eta d\phi$ (left) and average charged $p_{T}$ sum density, $dp^{sum}_{T} /d\eta d\phi$ (right), with $p_{T} >0.9 GeV/c$ and $|\eta| < 1$ versus the muon-pair invariant mass. Empty circles correspond to the raw (uncorrected) reconstruction level profiles; full circles correspond to the generator level profiles for the events passing the reconstruction level selection.}
\label{fig:cms_dy_ue}
\end{figure}

One can use the topological structure of hadron-hadron collisions to study the UE, only looking at the charged particles  produced in the interaction~\cite{cdf_ue}. Jets are constructed from the charged particles using a simple clustering algorithm. The direction of the leading charged particle jet is used to isolate regions of $\varphi$ space that are sensitive to the UE. 
The transverse region to the charged particle jet direction is almost perpendicular to the plane of the hard, back to back, scattering and is therefore very sensitive to the UE.
Typically the UE is studied looking at the charged track multiplicity and at $p_{T}^{sum}$ of tracks in the transverse region. 
These distributions as a function of leading jet $p_{T}$, ranging from 500 MeV/c or 900 MeV/c (depending on the lowest threshold for reconstruction), are quite interesting. 
The PYTHIA tunes predicts that these distribution rise quickly and then reach approximately a flat plateau at $p_{T} \sim$ 20 GeV/c.. This behavior  is a characteristic of the MPI. Then, at $p_{T} \sim$ 50 GeV/c, these distributions begin to rise again due to initial and final state radiation, which increases as the scale of the hard scattering. The HERWIG Montecarlo, where MPI is not implemented, has very different predictions.. The charged object distributions in jet topologies are studied by ATLAS and CMS with full simulation (Figures~\ref{fig:atlas_ue} and Figure~\ref{fig:cms_ue}). In the ATLAS studies (Figure~\ref{fig:atlas_ue}) the leading calorimetric jet is used as reference for the energy scale of the process and the investigated region is extended up to 1 TeV. Results are summarized in terms of the average number of reconstructed tracks in the transverse region and the average $p_{T}^{sum}$ of those tracks~\cite{moraes}. In the CMS studies (Figure~\ref{fig:cms_ue}) the energy scale is defined by the leading charged jet of the event and the observables in the transverse region are the average charged particles and $p_{T}^{sum}$ densities (left and right Figure~\ref{fig:cms_ue})\cite{noi}. Drell Yan muon pair production has been used by CMS to study the UE in an alternative and cleaner topology~\cite{noihold}. In these events the scale of the hard scattering is given by the muon pair invariant mass, while all the other charged tracks are attributed to the UE. 
Figure~\ref{fig:cms_dy_ue}  shows the prediction for the charged density and the energy density in the whole $\varphi$ region. 

\section{Conclusion}

Cross-sections and differential yields of charged particles produced in inelastic proton-proton collisions at $\sqrt{s} = 14$ TeV, can be measured with good precision with both CMS and ATLAS.

The predictions on the amount of hadronic activity in the region transverse to the jets produced 
in proton-proton interactions at the LHC energies are based on extrapolations from lower energy 
data (mostly from the Tevatron). These extrapolations are uncertain and predictions differ 
significantly among different model parameterizations. It is thus important to measure the UE activity at the LHC as soon as possible, and to compare those measurements with Tevatron data. 
This will lead to a better understanding of the QCD dynamics and to improvements of QCD 
based Monte Carlo models aimed at describing ``ordinary'' events at the LHC: an extremely important ingredient for ``new'' physics searches.

\section*{Acknowledgments}
Many people have contributed to the preparation of this talk with very high quality material and fruitful discussions: the CMS QCD group, Paolo Bartalini, Livio Fan\'o, Rick Field, Florian Bechtel, Klaus Rabbertz, Nikos Varelas, Ferenc Sikler, Richard Hollis, Aneta Iordanova, Fabiola Gianotti, Craig Buttar, Arthur Moraes, Iacopo Vivarelli.

\end{document}